\documentclass[]{IAC_style}

\usepackage{soul}
\usepackage[nolist,nohyperlinks]{acronym}
\usepackage{booktabs}
\usepackage{multirow}

\begin{document}

\newacro{iot}[IoT]{Internet of Things}
\newacro{nbiot}[NB-IoT]{Narrow-Band \ac{iot}}
\newacro{ntn}[NTN]{Non-Terrestrial Networks}
\newacro{geo}[GEO]{Geostrationary Orbit}
\newacro{leo}[LEO]{Low Earth Orbit}
\newacro{esa}[ESA]{European Space Agency}
\newacro{sdr}[SDR]{Software Defined Radio}
\newacro{dtsiot}[DtS IoT]{Direct-to-Satellite \ac{iot}}
\newacro{ttc}[TTC]{Telemetry and Telecommand}
\newacro{rf}[RF]{Radio-Frequency}

\IACpaperyear{2024} 
\IACpapernumber{IAC-24,B2,6,5,x86627} 
\IAClocation{Milan, Italy} 
\IACdate{14-18 October 2024} 

\IACcopyrightB{'copyright holder'}

\title{6GStarLab - A CubeSat Mission to support the development and standardization of Non-Terrestrial Networks towards 6G}

\IACauthor{Joan A. Ruiz-de-Azua$^{\orcidlink{0000-0001-5892-3404}}$}{1}{1}
\correspondingemail{joan.ruizdeazua@i2cat.net}

\IACauthor{Francesc Betorz$^{\orcidlink{0009-0004-4022-6626}}$}{1}{0}
\IACauthor{Hossein Rouzegar$^{\orcidlink{0000-0002-5517-0213}}$}{1}{0}
\IACauthor{Joan F. Munoz-Martin$^{\orcidlink{0000-0002-6441-6676}}$}{1}{0}
\IACauthor{Marc Badia$^{\orcidlink{0009-0002-2111-3461}}$}{2}{0}
\IACauthor{Roger Jove$^{\orcidlink{0000-0001-6447-9629}}$}{2}{0}
\IACauthor{Adriano Camps$^{\orcidlink{0000-0002-9514-4992}}$}{3}{0}
\IACauthor{Diego Lopez-Pizarro}{3}{0}
\IACauthor{Jordi Barrera}{4}{0}
\IACauthor{Jorge-Nicolas Alvarez $^{\orcidlink{0000-0001-6143-1515}}$}{4}{0}
\IACauthor{Ieremia Crisan}{4}{0}
\IACauthor{Mohammad Danesh$^{\orcidlink{0009-0003-7432-7137}}$}{5}{0}
\IACauthor{Vivek Mangalam$^{\orcidlink{0000-0002-4601-6767}}$}{5}{0}
\IACauthor{Jan Smisek$^{\orcidlink{0000-0003-0689-268X}}$}{5}{0}

\IACauthoraffiliation{Space Communications Research Group, i2CAT Foundation, Barcelona, Spain}

\IACauthoraffiliation{MWSE, Barcelona, Spain}

\IACauthoraffiliation{CommSensLab Department of Signal Theory and Communications, UPC BarcelonaTech, Barcelona, Spain}

\IACauthoraffiliation{Open Cosmos, Barcelona, Spain}

\IACauthoraffiliation{Transcelestial Technologies Pte Ltd, Singapore}

\abstract{
The emergence of the Non-Terrestrial Network (NTN) concept in the last years has revolutionized the space industry. This novel network architecture composed of aircraft and spacecraft is currently being standardized by the 3GPP. This standardization process follows dedicated phases in which experimentation of the technology is needed. Although some missions have been conducted to demonstrate specific and service-centric technologies, a open flexible in-orbit infrastructure is demanded to support this standardization process. This work presents the 6GStarLab mission, which aims to address this gap. Specifically, this mission envisions to provide a 6U CubeSat as the main in-orbit infrastructure in which multiple technology validations can be uploaded. The concept of this mission is depicted. Additionally, this work presents the details of the satellite platform and the payload. This last one is designed to enable the experimentation in multiple radio-frequency bands (i.e. UHF, S-, X-, and Ka-bands) and an optical terminal. The launch of the satellite is scheduled for Q2 2025, and it will contribute to the standardization of future NTN architectures.   
}

\maketitle
\thispagestyle{fancy} 


\section{Introduction}
In the last years, the space sector has experienced a novel revolution driven by the integration of mobile technologies. The emergence of \ac{ntn} \cite{azari2022evolution} has redefined the perception of satellite systems to provide ubiquitous and global connectivity for multiple broadband and \ac{iot} services. The standardization effort done by the 3GPP has resulted in the definition of \ac{ntn} as networks composed of spacecraft and aircraft that are integrated with terrestrial nodes, achieving a 3D-network architecture. The standardization process of \ac{ntn} is at a research phase, where key players are open to collaborate to assess the network feasibility and interoperability \cite{larsson2024}. This phase is characterized by a strong trial and error interaction towards continuous improvements. For this reason, realistic and open experimental infrastructure is needed to support this phase. 

Some previous experimentation has been conducted in a laboratory-controlled environment \cite{kodheli2021random, kellermann2022novel}. Although necessary to verify the development, this kind of environment is not representative enough to support the standardization process. Specifically, performance of novel protocols and developments has to be measured to provide the required insights that support the details of the standardization. For this reason, dedicated satellite missions have been conducted in the last years. 

Private companies have followed a strategic plan to demonstrate the capabilities of satellite systems to provide 5G connectivity. Authors in \cite{mediatek2020} demonstrated a \ac{nbiot} connection between a sensor device and a \ac{geo} satellite. Although this large-delay connection is suitable for \ac{iot} traffic, the fixed coverage may limit the service provision. For this reason, another initiative is validating its \ac{nbiot} service with \ac{leo} satellites \cite{sateliot2024}. Technology in-orbit demonstrations are also conducted concerning the provision of broadband services. Authors in \cite{ast2023} have validated the a direct-to-handset connection between a \ac{leo} satellite and a standard mobile terminal. These are just few examples of how private ecosystem is pushing towards the in-orbit demonstration of their technologies. Although these initiatives clearly contributes to the standardization process, they target specific developments rather than large technology exploration and at the same time they may not be affordable for all entities. For this reason, open and flexible experimental in-orbit infrastructure is still needed.

The \ac{esa} has identified this gap, and they conceived a dedicated missions to provide experimental infrastructure to support 3GPP standardization \cite{esa2023}. This mission, called \textit{Sterling}, envisions to provide an open in-orbit infrastructure with a satellite system that enables to upload experiments that leverage on its hardware capabilities, like Ka-/W-bands and optical interfaces. Another initiative, called \textit{SeRANIS} \cite{bachmann2024seranis} follows the same principle of the \ac{esa} one, but it presents the continuous deployment of micro-satellites to ensure in-orbit experimental infrastructure over the years. Although all these approaches provides large experimental capabilities, they are directly driven by (1) the missions timeline, and (2) the cost of their missions. The associated missions are envisioned to be conducted in the following 3-5 years, which place them misaligned with the standardization phases in which performance results are needed. Additionally, the cost of the infrastructure may limit its access. An alternative approach has been considered to address these limitations.

This work contributes thus with the presentation of the 6GStarLab mission concept and the details of the satellite system. Specifically, the main contributions are (1) the details of the mission (e.g. the access to this infrastructure or the service areas); (2) the design of the satellite platform, highlighting the interfaces and the components characteristics; (3) the design of the multi-front-end flexible payload; and (3) the following steps of the mission that will be conducted. 

The remainder of the work is structured as follows. Section \ref{sec:mission} presents the concept of the 6GStarLab mission. The details of the satellite platform and the embedded payload are presented in Sections \ref{sec:satellite} and \ref{sec:payload} respectively. Finally, Section \ref{sec:conclusions} concludes the work.

\section{6GStarLab mission concept}
\label{sec:mission}

\begin{figure}
    \centering
    \includegraphics[width=\columnwidth]{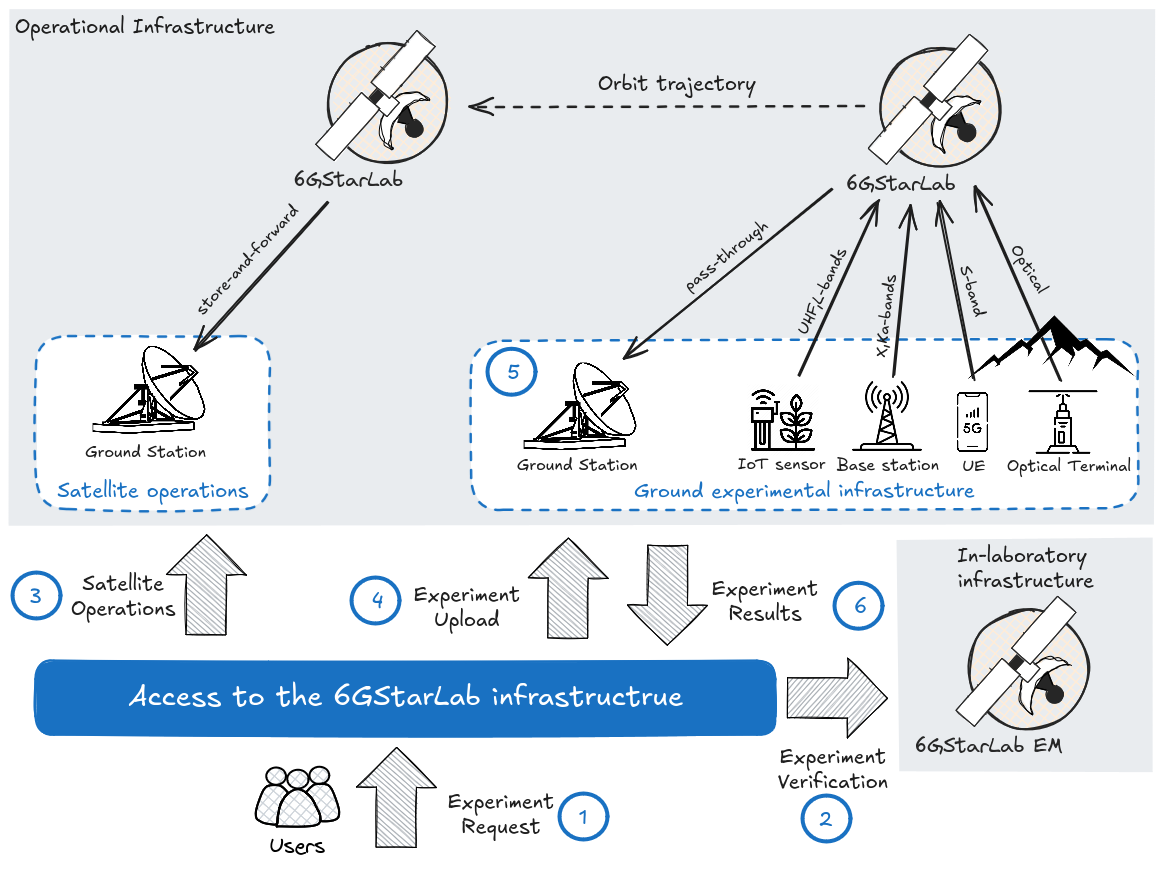}
    \caption{6GStarLab mission concept overview}
    \label{fig:mission}
\end{figure}

The 6GStarLab mission aims to establish an open in-orbit infrastructure that fosters the development and standardization of 6G \ac{ntn}. As illustrated in Figure \ref{fig:mission}, the mission seeks to deploy an open, in-orbit laboratory that facilitates the experimentation of innovative solutions for the integration of space and terrestrial 6G networks. Unlike the previous missions, this one is composed of a \ac{leo} 6U CubeSat that provides the capabilities to experiment with these novel technologies. Although its performance (e.g. throughput) may be limited by the platform resources, its design provides flexibility in the frequencies, terminals and on-board processing architecture and components. The end-users of this laboratory will be able to upload the necessary software to execute their experiments leveraging on their hardware capabilities. For this reason, the satellite is designed following the virtualization principle, which ensures a fully programmable capabilities. This features are achieved by the integration of the Flexible Payload framework, presented in previous works \cite{montilla2024, ntontin2024ether}. Additionally, the satellite includes the necessary hardware to satisfy this flexibility with two \ac{sdr} and a set of radio front-ends working in the following bands: UHF, S-band, X-band, and Ka-band. Additionally, an optical terminal is included to support the research and developments on optical networks. Although other missions may provide better performance in these bands, the possibility of having these multiple front-ends enables to experiment with all these bands. This flexibility has predominated in the design with respect the maximum performance.

\section{Satellite platform}
\label{sec:satellite}
\begin{figure}
    \centering
    \includegraphics[width=\linewidth]{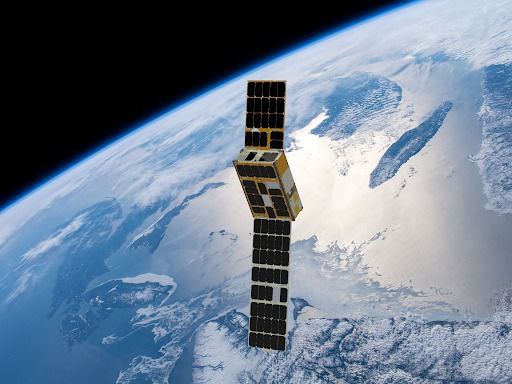}
    \caption{Preliminary overview of the satellite platform from Open Cosmos}
    \label{fig:cubesat}
\end{figure}

The 6GStarLab mission is based on a 6U CubeSat platform (Figure \ref{fig:cubesat}) designed to support the required flexibility to conduct a wide range of experiments. The platform is engineered to deliver reliable performance in space, making it suitable for dedicated experimental missions. This platform includes thus the essential subsystems to ensure the seamless execution of the experiment in the targeted area. Table \ref{tab:specs} presents the main characteristics of the platform, which is developed by Open Cosmos company. The platform is designed to ensure 15\% of payload execution duty cycle, and achieve three years of lifetime.

\begin{table}
    \centering
        \caption{Specifications of the 6GStarLab platform}
    \begin{tabular}{l p{0.21\textwidth}}
         \toprule
         Specification & Value \\
         \midrule
         \multicolumn{2}{c}{\textit{Volume and mass capacity}} \\
         \midrule
         Form Factor & 6U CubeSat \\
         Payload mass & $\le 8$ kg \\
         Payload volume & $\le 4$ U \\
         \midrule
         \multicolumn{2}{c}{\textit{Power capabilities}}  \\
         \midrule
         Available Average power & 7-20 W \\
         Payload peak power & 160 W \\
         Battery capacity & 42-120 Wh \\
         \midrule
         \multicolumn{2}{c}{\textit{Data Handling}}  \\
         \midrule
         Data buses & I2C, UART, SPI, Spacewire, Ethernet, CAN \\
         Data storage & 200 GB \\
         \midrule
         \multicolumn{2}{c}{\textit{Communication}}  \\
         \midrule
         Operations & S-band up to 5 Mbps \\
         \bottomrule
    \end{tabular}
    \label{tab:specs}
\end{table}

\section{Satellite Payload}
\label{sec:payload}

\subsection{System overview}
The 6GStarLab payload is designed with a modular architecture to provide high
flexibility and enable advanced experimentation in both \ac{rf} and optical
communications. At its core, the \ac{rf} payload consists of two Software Defined Radios
(SDRs), known as Minerva, each equipped with two front-ends and its corresponding
antenna connectors. Each SDR occupies 1U of CubeSat volume and interfaces with
the satellite platform via Ethernet (for data) and UART (for telemetry and commands).
Managed by the Flexible Payload Framework developed by Fundació i2CAT, the SDR
configuration offers wide flexibility by supporting real-time onboard reconfiguration,
allowing researchers to implement and run experiments on demand.

Both SDRs can operate simultaneously, enabling multiband operations across several
frequency bands, including UHF, S-band, X-band, and Ka-band. They are
interconnected via a high-speed interface to enable fast data exchange and to provide
redundancy in data storage. This dual SDR setup enhances the system ability to
support complex experiments across multiple bands while ensuring robust data
handling and operational flexibility.

In addition to its \ac{rf} capabilities, the 6GStarLab payload includes an optical
communication terminal for high-speed data transmission through a laser-based space-
to-ground link. The optical payload also occupies 1U of CubeSat volume and interfaces
with the satellite platform via Ethernet (for data) and CAN (for telemetry and
commands). During optical operations, the laser terminal’s pointing system works in
coordination with the satellite’s attitude control system to ensure the required precise
alignment between the onboard laser device and the ground station. All three payloads
are interconnected through a dedicated Ethernet bus managed by the On-Board
Computer (OBC), allowing continuous data interchange among the various payload
subsystems.

The duty cycle of the payload is at least $15\%$ of the satellite’s orbital period. During this time, the platform continuously and uninterruptedly provides all necessary services to the payload, including power, communications, and precise pointing capabilities. The power requirements for the payloads are as follows: each SDR in standby mode
consumes up to 10 W, with a maximum active consumption of 30 W, while the optical
laser requires 4 W in standby and a maximum of 25 W when operational. The platform
can supply a continuous nominal power output of 45 W throughout the entire duty cycle
and can deliver a total maximum power of 85 W.

By combining the flexibility of SDR-based multiband operations with advanced optical
communication capabilities, the 6GStarLab payload represents a unique in-orbit
laboratory designed to significantly advance space communications research and
support the standardization of future 6G NTN technologies.

\begin{table*}[]
    \centering
    \caption{Front-ends of the 6GStarLab payload and its corresponding frequency bands}
    \begin{tabular}{c c c c}
        \toprule
         Front-ends & Potential use case  & Frequency band & Frequency range \\
         \midrule
         \multirow{3}{*}{Front-end \#1} & \multirow{3}{*}{\ac{dtsiot} \cite{fraire2019direct}} & \multirow{3}{*}{UHF} & 433.00 MHz to 434.79 MHz \\
         &  & & 863.00 MHz to 870.00 MHz \\
         &  & & 903.00 MHz to 914.20 MHz \\
         \midrule
         \multirow{2}{*}{Front-end \#2} & \multirow{2}{*}{\ac{dtsiot} NTN (n256) \cite{3gpp_rel17}} & L/S-band & 1980.00 MHz to 2025.00 MHz \\
         & & S-band & 2160.00 MHz MHz to 2200.00 MHz \\
         \midrule
         Front-end \#3 & Data backhauling & X-band & 10.45 GHz to 10.50 GHz \\
         \midrule
         \multirow{2}{*}{Front-end \#4} & \multirow{2}{*}{NTN (n511) \cite{3gpp_rel17}} & \multirow{2}{*}{Ka-band} & 19.30 GHz to 20.10 GHz \\
         & & & 29.10 GHz to 30.00 GHz \\
         \bottomrule
    \end{tabular}
    \label{tab:frontends}
\end{table*}

\subsection{Flexible payload software framework}

As a software-defined satellite platform, flexible payload refers to a satellite payload whose features and functionalities are reconfigurable in real time via software updates and modifications. This adaptability lowers costs by enabling the satellite to adjust to varying mission requirements and user demands without requiring physical hardware modifications \cite{marziale2006flexible,paillassa2003flexible}. As 6GStarlab is envisioned to test various services on-demand in a different scenarios, a flexible payload with aforementioned capabilities could be a solution. 

Typically, when designing a conventional payload, software and electronic parts are selected with specific functionalities. This leads to a strong reliance of services on platform resources which restricts the payload's ability to for only certain missions. \ac{sdr} is suggested as a method to get around these limitations by utilizing programmable hardware or setting up procedures to update the software in order to implement radio functions like modulation/demodulation or signal processing. \ac{sdr} allows for adaptability to new communication protocols and channel assignment policies by modifying the digital hardware and leveraging existing analog front-ends \cite{nguyen2020communication}. This technique creates an independence between services and the platform enabling the satellites to reallocate their power, frequency, time, or coverage resources in response to changing conditions. Furthermore, it mitigates the risks by allowing for the mission to be reoriented through defining new tasks in the event that the original mission fails or is not market beneficial \cite{vidal2021methodology,balty2007flexible}.

\subsection{Radio interfaces}
6GStarLab is designed to support operations across multiple spectral bands, providing
enhanced flexibility and making it a unique infrastructure for researchers in telecommunications. This multi-frequency capability will enable research in the UHF, L,
S, X, and Ka bands, opening the door to a wide range of applications. These include,
for example, research on advanced 5G and 6G technologies using \ac{ntn} and their integration with terrestrial infrastructure, and the
advancement and research of Internet of Things technologies from space, such as
LoRa and NB-IoT. This versatility makes 6GStarLab an invaluable platform for
advanced research in space communications and space-based IoT applications.

6GStarLab integrates two \ac{sdr}s and a set of radio front-
ends, enabling operations in the previously mentioned frequency bands. Table 1
provides a summary of the frequency bands and the specific ranges covered,
highlighting the satellite’s versatility and broad frequency support.

The radio interfaces and link configurations will enable the following capabilities: an S-
band gross data rate up to 1.152 Mbps (144 kBps) with a maximum bandwidth of 0.75 MHz, an X-band gross data rate
up to 2.3 Mbps (288 kBps) with a maximum bandwidth of 1.5 MHz, and a Ka-band gross data rate up to 4.6 Mbps (576
KBps) with a bandwidth of 3 MHz. Both \ac{sdr}s, along with their corresponding front-ends
can operate simultaneously enabling transmissions at different frequencies in parallel.
This capability will allow for multiple configurations, providing researchers with greater
flexibility and the ability to conduct more complex communication experiments across
various spectral bands.

\subsection{Optical terminal}
6GStarLab will also provide advanced infrastructure and equipment to conduct
research in laser-based optical communications. A dedicated on-board laser terminal,
integrated into a compact 1U payload, will enable experimentation with high-speed
data transmission through space-to-ground laser links.

The laser terminal will consist of several key components: an optical telescope, a high-
resolution infrared camera, two collimated infrared laser light sources for data
transmission, an Erbium Yterbium Doped Fiber Amplifier (EYDFA) to boost the laser signal strength,
and the necessary electronics for processing, control, and stabilization.

This Laser Terminal (see technical specifications in Table \ref{tab:laser_terminal_specs}) is designed to offer a
high-speed, reliable Space-to-Ground data downlink system, utilizing both beacon-based and beaconless
pointing and accurate acquisition and tracking mechanisms. These systems have been
optimized to be fully compatible with the satellite’s pointing capabilities, ensuring stable and precise alignment between the satellite and the optical ground station.

\begin{table}[]
    \caption{Hardware specifications of the Laser Terminal}
    \centering
    \begin{tabular}{cc}
        \toprule
        \textbf{Specification} & \textbf{Value} \\
        \midrule
        \multirow{2}{*}{Mass and Size} & 950 g \\
                                       & 1U (97x97x90 mm) \\
        \midrule
        \multirow{2}{*}{Electrical Power} & 4 W average \\
                                          & 25 W peak \\
        \midrule
        Data Interface & 1Gbit Ethernet, serial \\
        \midrule
        \multirow{2}{*}{Downlink (Space to Earth)} & 1 W at 1530 nm \\
                                                   & up to 1 Gbps \\
        \midrule
        \multirow{2}{*}{Uplink (Earth to Space)} & 1560 nm \\
                                                 & 100 Mbps \\
        \midrule
        Distance range & 500 to 1500 km \\
        \midrule
        Required pointing & 1º (3$\sigma$) \\
        \bottomrule
    \end{tabular}
    \label{tab:laser_terminal_specs}
\end{table}

The laser communication technology within 6GStarLab will enable research in high-
speed downlink and uplink data transmission. It will also facilitate advanced research
into reliable and faster space-to-Earth and intersatellite pointing mechanisms.
The laser terminal is designed to establish communication from orbit when in direct
line-of-sight with a dedicated optical ground station, which will also be provided as part
of the infrastructure supporting 6GStarLab mission. This ground station will play a
crucial role in the validation and optimization of the laser communication system,
enabling real-time data transfer and performance analysis.

By integrating this compact laser communication system, 6GStarLab aims to push the
boundaries of laser-based space communication technologies in CubeSats, fostering
innovation in both terrestrial and extraterrestrial data links.

\subsection{Hardware modularity}
\begin{figure*}
    \centering
    \includegraphics[width=\textwidth]{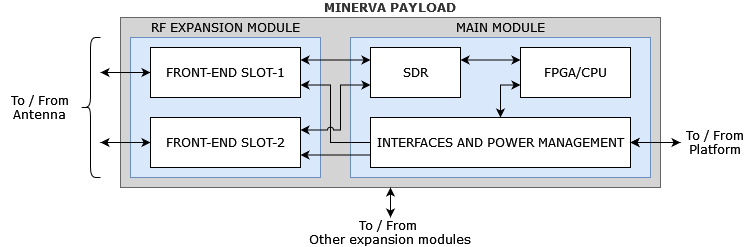}
    \caption{Hardware architecture of Minerva payload in the 6GStarLab mission}
    \label{fig:HW}
\end{figure*}

To build the flexible payload, a multi front-end \ac{sdr} platform, called \textit{Minerva} payload, is used. Minerva is based on a modular architecture. As Figure \ref{fig:HW} depicts, the architecture is based by a main module and an \ac{rf} expansion module. The main module is responsible to manage power, interfaces and process information. It includes a \ac{sdr}, a power and interfaces circuitry and a based FPGA/CPU control unit. Then, \ac{rf} expansion module contains two available slots for front-ends. 

To achieve all the flexible payload frequency capabilities, two Minerva payloads are used. The front-end slots are occupied by (1) UHF, (2) S-Band, (3) X-Band, and (4) Ka-Band front-ends. Each Minerva payload is powered and controlled independently by the platform, but contains high-speed data links for data sharing (if necessary). 

\subsection{Antenna subsystem}
Table \ref{tab:frontends} above summarizes the four front-ends, use cases, frequency bands, and ranges of frequencies to be covered. The space available for all the different antennas is just a 2 x 2 U$^2$ in the 3 x 2 U$^2$ side pointing to nadir, and the maximum antenna height is determined by the space between the CubeSat and the internal walls of the CubeSat deployer (e.g. 25 mm in an Exopod NOVA \cite{exopod}). The volume of the two tuna cans (82.0 mm diameter, 58.0 mm depth) is also available. 
Since link budget requirements led to antenna gains of 10 dB at S-band, and 12 dB at higher bands, given the above constraints, a preliminary design indicates that the UHF antennas will be a 2 x 1 array of dual-band charged monopoles located in the tuna cans. The L-, S-, X- and Ka-band antennas will be 2 x 2 patch arrays, with gains at the higher bands slightly lower than the 12 dB requirement, but large enough to safely close the link. The L- and S-band arrays will be stacked one on top of the other, while the X-, and Ka-band arrays -because of their smaller size- will be stacked on top of an S-band patch each. 

\section{Conclusions and Future steps}
\label{sec:conclusions}
In the last years, the space sector has experienced a novel revolution driven by the integration of mobile technologies. The emergence of \ac{ntn} \cite{azari2022evolution} has redefined the perception of satellite systems to provide ubiquitous and global connectivity. Although huge efforts are being conducted in different research activities, the deployment of NTN in the future is driven by the available experimental infrastructure. Specifically, dedicated in-orbit assets are needed to provide empirical performance and enable technology validation towards the standardization of this novel concept. Although some private service-centered initiatives have been conducted, an experimental open in-orbit infrastructure is still a gap that needs to be addressed. 

This work has presented an overview of the 6GStarLab mission. This mission envisions to provide an open in-orbit infrastructure that supports the novel developments towards the standardization and deployment of 6G \ac{ntn}. Thanks to the deployment of a 6U CubeSat and a flexible payload approach, the users of this infrastructure will be able to upload dedicated experimental software that leverages the satellite hardware capabilities, like multi-radio front-ends and optical communications. Details about the mission concept and the design of the satellite and its payload have been presented. Its radio capabilities have been discussed, remarking the flexibility in multiple bands (e.g. UHF, S-, X-, and Ka-bands) and the combination with an optical terminal. Additionally, the design includes a software framework, called Flexible Payload, that provides a virtualization layer that simplifies the installation of external software, and thus facilitates the execution of the experiments. 

Additionally, a preliminary view of the procedure to access this infrastructure has been presented, highlighting the need to satisfactorily pass an in-laboratory verification with a FlatSat model. With this procedure, the users will be able to verify its experiments before tackling the in-orbit infrastructure. With all these features, the 6GStarLab mission envisions to provide the current needed experimental infrastructure to support future developments, and the standardization of the future NTN concept. Currently, the mission is in the implementation phase, and it is expected that the satellite will be launched in Q2 2025. 

\section*{Acknowledgements}
This work has been co-funded by the Grant TSI-064100-2023-18 funded by the Ministry of Economic Affairs and Digital Transformation and by the “European Union NextGenerationEU/PRTR” within the call "Ayudas UNICO I+D 6G 2023 (Infraestructuras de investigación y equipamiento científico-técnico)”, and by the Spanish Ministry of Economic Affairs and Digital Transformation and the European Union – NextGeneration EU, in the framework of the Recovery Plan, Transformation and Resilience (PRTR) (Call UNICO I+D 5G 2021, ref. number TSI-063000-2021-5-6GSatNet-SS, and TSI-063000-2021-8-6GSatNet-SeS).

\bibliographystyle{unsrt}
\bibliography{biblio}

\begin{thebibliography}{10}

\bibitem{azari2022evolution}
M~Mahdi Azari, Sourabh Solanki, Symeon Chatzinotas, Oltjon Kodheli, Hazem Sallouha, Achiel Colpaert, Jesus Fabian~Mendoza Montoya, Sofie Pollin, Alireza Haqiqatnejad, Arsham Mostaani, et~al.
\newblock Evolution of non-terrestrial networks from 5g to 6g: A survey.
\newblock {\em IEEE communications surveys \& tutorials}, 24(4):2633--2672, 2022.

\bibitem{larsson2024}
Daniel Chen~Larsson, Asbjörn Grövlen, Stefan Parkvall, and Olof Liberg.
\newblock 6g standardization – an overview of timeline and high-level technology principles.
\newblock {\em Ericsson Blog}, 2024.

\bibitem{kodheli2021random}
Oltjon Kodheli, Abdelrahman Astro, Jorge Querol, Mohammad Gholamian, Sumit Kumar, Nicola Maturo, and Symeon Chatzinotas.
\newblock Random access procedure over non-terrestrial networks: From theory to practice.
\newblock {\em IEEE Access}, 9:109130--109143, 2021.

\bibitem{kellermann2022novel}
Timo Kellermann, Roger~Pueyo Centelles, Daniel Camps-Mur, Ramon Ferr{\'u}s, Marco Guadalupi, and Anna~Calveras Aug{\'e}.
\newblock Novel architecture for cellular iot in future non-terrestrial networks: store and forward adaptations for enabling discontinuous feeder link operation.
\newblock {\em IEEE access}, 10:68922--68936, 2022.

\bibitem{mediatek2020}
Dan Oliver.
\newblock Mediatek and inmarsat deliver world’s first 5g nb-iot test via geo satellite.
\newblock {\em 5GRadar}, August 2020.

\bibitem{sateliot2024}
IoTNow team.
\newblock Sateliot launches four satellites to enhance global iot connectivity.
\newblock {\em IoTNow}, August 2024.

\bibitem{ast2023}
SpaceRef.
\newblock Ast spacemobile achieves space-based 5g cellular broadband connectivity from everyday smartphones.
\newblock {\em Spacenews}, September 2023.

\bibitem{esa2023}
Antonio Franchi, Xavier Lobao, and Stefano Cioni.
\newblock {6G Satellite Precursor – Open In-Orbit 6G Laboratory}, 2023.
\newblock Avilable online at: \url{http://www-groups.dcs.st-and.ac.uk/~history/Biographies/Noether_Emmy.html}, last accessed on September 10th, 2024.

\bibitem{bachmann2024seranis}
Johannes Bachmann, Artur Kinzel, Francesco Porcelli, Alexander Schmidt, Robert Schwarz, Christian Hofmann, Roger F{\"o}rstner, and Andreas Knopp.
\newblock {\em SeRANIS: In-Orbit-Demonstration von Spitzentechnologie auf einem Kleinsatelliten}.
\newblock Deutsche Gesellschaft f{\"u}r Luft-und Raumfahrt-Lilienthal-Oberth eV, 2024.

\bibitem{montilla2024}
Victor Montilla, Alessandro Villegas, and Joan~A. Ruiz-de Azua.
\newblock The flexible payload - a virtualization framework to autonomously deploy software-based payloads on fpga soc platforms.
\newblock In {\em Small Satellites \& Services International Forum (SSSIF)}. IEEE, 2024.

\bibitem{ntontin2024ether}
K~Ntontin, L~Tomaszewski, JA~Ruiz-de Azua, A~C{\'a}rdenas, R~Pueyo Centelles, C-K Lin, A~Mesodiakaki, A~Antonopoulos, N~Pappas, M~Fiore, et~al.
\newblock Ether: A 6g architectural framework for 3d multi-layered networks.
\newblock In {\em 2024 IEEE Wireless Communications and Networking Conference (WCNC)}, pages 1--6. IEEE, 2024.

\bibitem{fraire2019direct}
Juan~A Fraire, Sandra C{\'e}spedes, and Nicola Accettura.
\newblock Direct-to-satellite iot-a survey of the state of the art and future research perspectives: Backhauling the iot through leo satellites.
\newblock In {\em International Conference on Ad-Hoc Networks and Wireless}, pages 241--258. Springer, 2019.

\bibitem{3gpp_rel17}
{Technical Specification Group Radio Access Network}.
\newblock {\em User Equipment (UE) radio transmission and reception; Part 5: Satellite access Radio Frequency (RF) and performance requirements (Release 18)}.
\newblock 3rd Generation Partnership Project (3GPP), June 2024.
\newblock V18.6.0.

\bibitem{marziale2006flexible}
Vincenzo Marziale, Alessandro Pisano, and Giampiero Di~Paolo.
\newblock Flexible payload technologies to enable multi mission satellite communication systems.
\newblock In {\em 24th AIAA International Communications Satellite Systems Conference}, page 5374, 2006.

\bibitem{paillassa2003flexible}
B{\'e}atrice Paillassa and Catherine Morlet.
\newblock Flexible satellites: software radio in the sky.
\newblock In {\em 10th International Conference on Telecommunications, 2003. ICT 2003.}, volume~2, pages 1596--1600. IEEE, 2003.

\bibitem{nguyen2020communication}
Hung~H Nguyen and Peter~S Nguyen.
\newblock Communication subsystems for satellite design.
\newblock In {\em Satellite Systems-Design, Modeling, Simulation and Analysis}. IntechOpen, 2020.

\bibitem{vidal2021methodology}
Florian Vidal, Herv{\'e} Legay, George Goussetis, Maria Garcia~Vigueras, S{\'e}gol{\`e}ne Tubau, and Jean-Didier Gayrard.
\newblock A methodology to benchmark flexible payload architectures in a megaconstellation use case.
\newblock {\em International Journal of Satellite Communications and Networking}, 39(1):29--46, 2021.

\bibitem{balty2007flexible}
C{\'e}dric Balty and Jean-Didier Gayrard.
\newblock Flexible satellites: A new challenge for the communication satellite industry.
\newblock In {\em 25th AIAA International Communications Satellite Systems Conference (organized by APSCC)}, page 3256, 2007.

\bibitem{exopod}
EXOpod.
\newblock {EXOpod NOVA User Manual}, 2024.
\newblock Available online at: \url{https://exolaunch.com/documents/EXOpod_Nova_User_Manual_March_2024.pdf}, last accessed on September 17th, 2024.

\end{thebibliography}

\end{document}